\documentclass{ws-procs975x65}            
\usepackage{hyperref} 
\begin{document}
\hyphenation{ALICE}
\title{Diffraction physics with ALICE\\
at the LHC}

\author{Sergey Evdokimov for the ALICE collaboration}

\address{Institute for High Energy Physics of NRC "Kurchatov Institute",\\
Protvino, 142281, Russia\\
E-mail: Sergey.Evdokimov@cern.ch}

\begin{abstract}
The ALICE experiment is equipped with a wide range of detectors providing excellent tracking 
and particle identification in the central region, as well as forward detectors with extended 
pseudorapidity coverage, which are well suited for studying diffractive processes. 
Cross section measurements of single and double diffractive processes 
performed by ALICE in pp collisions at $\sqrt{s}=0.9,~2.76,~7$~TeV will be reported. Currently, ALICE is studying double-gap events in pp collisions at $\sqrt{s}=7$~TeV, which give an insight into the central 
diffraction processes: current status and future perspectives will be discussed. The upgrade plans for diffraction studies, 
further extending the pseudorapidity acceptance of the ALICE setup for the forthcoming Run 2 
of the LHC, will be outlined.

\end{abstract}


\bodymatter

\section{Introduction}\label{aba:sec1}
The total proton-proton cross section receives contributions from different processes. There are significant contributions from elastic scattering ($\sim$ 26\%), single ($\sim$ 12.5\%), double ($\sim$ 6.5\%) and central ($<$1\%) diffraction at the LHC energies\cite{XSfold}. These processes usually acquires at low $t$-values and are subjects of non-perturbative QCD. They also could be described in terms of the Regge theory in which proton-proton scattering is interpreted as a Reggeon exchange in $t$-channel of the reaction. The Regge pole approximation suggests a power-low growth of the total cross section with a squared collision energy $s$: 
\begin{equation}
\sigma_{\rm tot} \sim {\left( \frac{s} {s_0}\right)}^{\alpha(0) - 1},
\label{aba:eq1}
\end{equation}
where $\alpha(t)$ is the Reggeon trajectory. Experimental studies have confirmed the cross section growth with the collision energy, known as the Serpukhov effect\cite{SerpEff} and confirmed to be universal for all hadrons at CERN\cite{ppGrowthCERNISR,ppGrowthCERNSppS} and Fermilab\cite{ppGrowthFermilab}. To describe this effect, V.Gribov introduced\cite{gribov} the so-called supercritical Pomeron trajectory with the intercept $\alpha(0) > 1$. All trajectories associated with known particles have intercepts $\alpha(0) < 1$ and their contributions to the total cross section become negligible at high energies\cite{levin}. Therefore, the Pomeron exchange dominates at high energies. In this way, diffraction studies help in understanding the Pomeron nature and its connection to the soft QCD processes and vice versa.
From the experimental point of view, diffractive processes are characterised by the presence of large gaps in particle rapidities. A mean rapidity distance between charged particles in non-diffractive events is about $\Delta y \sim 0.2$ at the LHC energies. In diffraction processes, protons can break up into clusters with proton quantum numbers except of parity and spin. These clusters can be separated by a rapidity interval called a large rapidity gap. Processes studied at ALICE are as follows:
\begin{itemlist}
\item Non-diffractive (ND) -- no large rapidity gaps for the secondary particles;
\smallskip
\item Single diffraction (SD) -- large rapidity gap between a particle cluster and one of the outgoing protons;
\smallskip
\item Double diffraction (DD) -- large central rapidity gap between two clusters;
\smallskip
\item Central diffraction (CD) -- a cluster in the central region separated from outgoing protons by two large rapidity gaps.
\end{itemlist}

\section{Experimental setup}
The ALICE detector\cite{alice} consists of a large number of detector subsystems, and has a unique potential for diffractive physics in terms of excellent tracking and particle identification (PID) capabilities as well as extended pseudorapidity coverage. The central barrel consists of the Inner Tracking System (ITS), the Time Projection Chamber (TPC), the Transition Radiation Detector (TRD), the Time Of Flight detector (TOF), the High Momentum Particle Identification Detector (HMPID) and two electromagnetic calorimeters (PHOS and EMCAL). The set of the central barrel detectors provides excellent particle tracking and identification capabilities in the pseudorapidity range $| \eta | < 0.9$. The ITS has two layers of silicon pixel detectors covering the pseudorapidity range $| \eta | < 2 $ and $| \eta | < 1.4 $ for the inner and outer layers respectively. Fig.\ref{fig:fig1} shows the performance of the PID in ALICE: the very good low-momentum particle identification plays a key role for studying central diffraction in various channels ($\pi\pi$, $KK$, ...). Forward detectors allow us to trigger on events with large rapidity gaps by the absence of any signals in them. The VZERO detector consists of two scintillator arrays with the pseudorapidity ranges $ -3.7 < \eta < -1.7 $ and $ 2.7 < \eta < 5.1$, respectively. The Forward Multiplicity Detector (FMD) is designed to measure charged particles in pseudorapidity ranges $ -3.4 < \eta < -1.7 $ and $ 1.7 < \eta < 5.0$ and consists of five rings of silicon semiconductor detectors. Thus the effective pseudorapidity coverage of the ALICE detectors for diffractive studies is $ -3.7 < \eta < 5.1 $ which is almost 9 units.

\def\figsubcap#1{\par\noindent\centering\footnotesize(#1)}
\begin{figure}[h]%
\begin{center}
  \parbox{2.1in}{\includegraphics[width=2in]{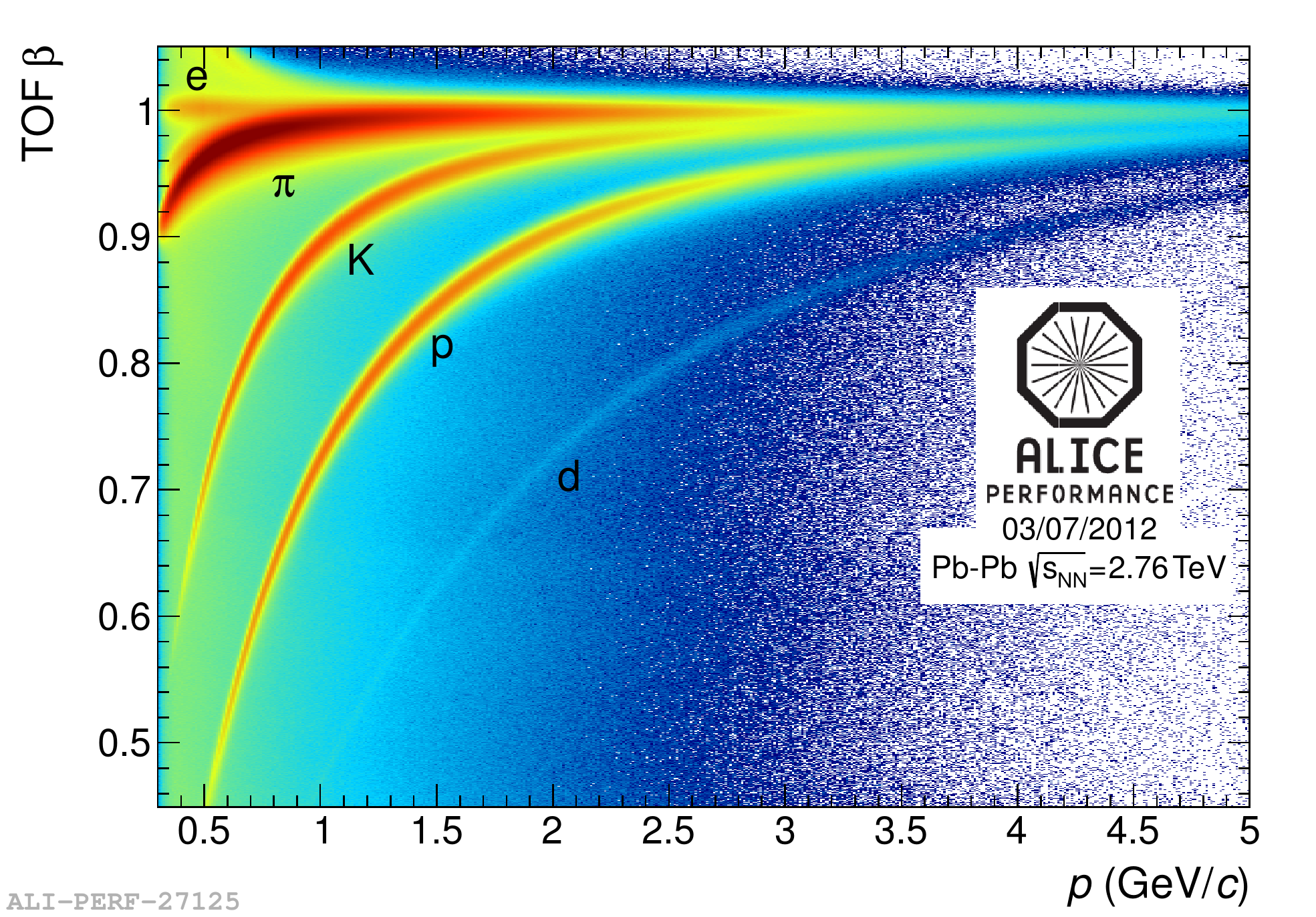}\figsubcap{a}}
  \hspace*{4pt}
  \parbox{2.1in}{\includegraphics[width=2in]{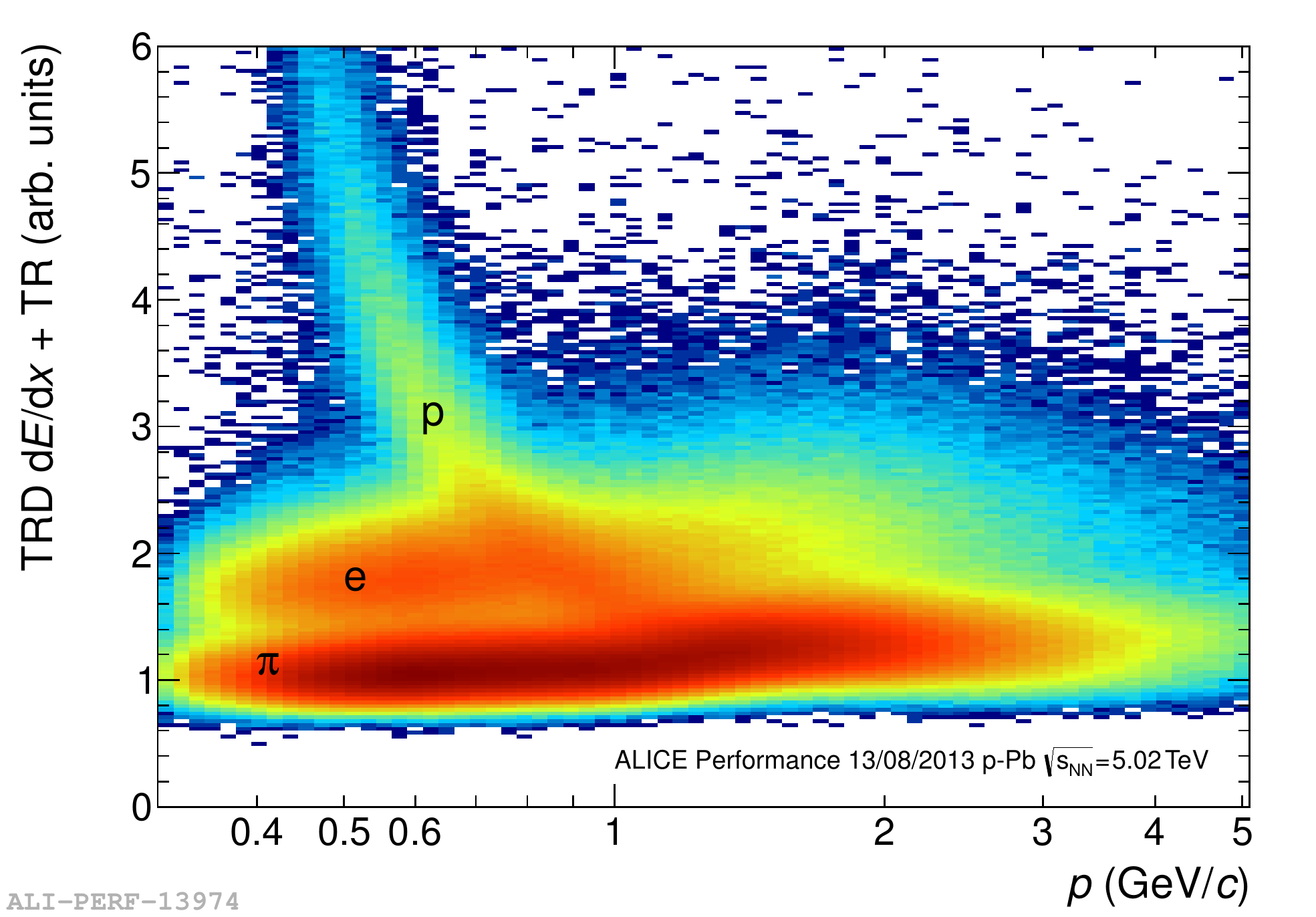}\figsubcap{b}}
  
  \parbox{2.1in}{\includegraphics[width=2in]{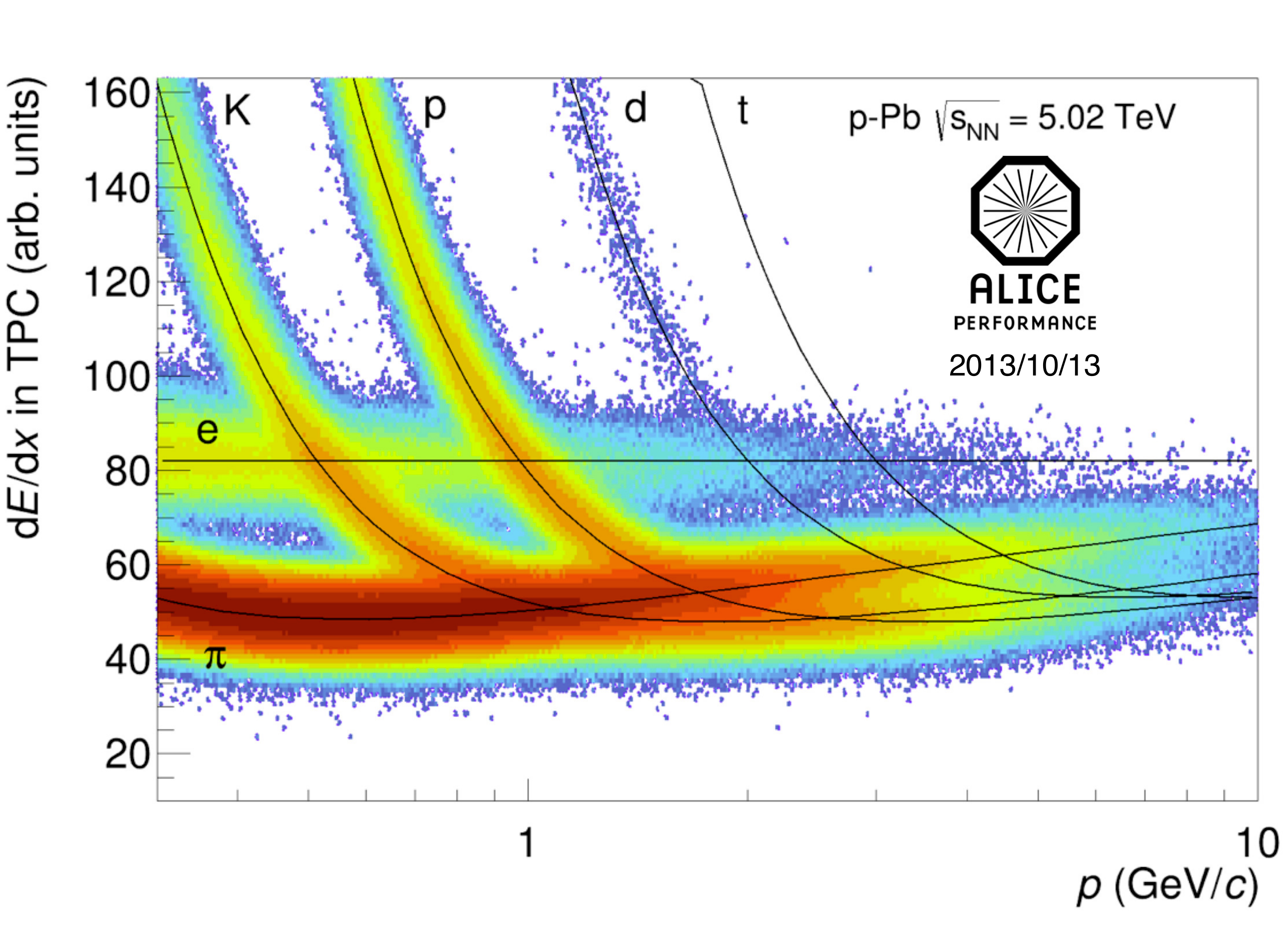}\figsubcap{c}}
  \hspace*{4pt}
  \parbox{2.1in}{\includegraphics[width=2in]{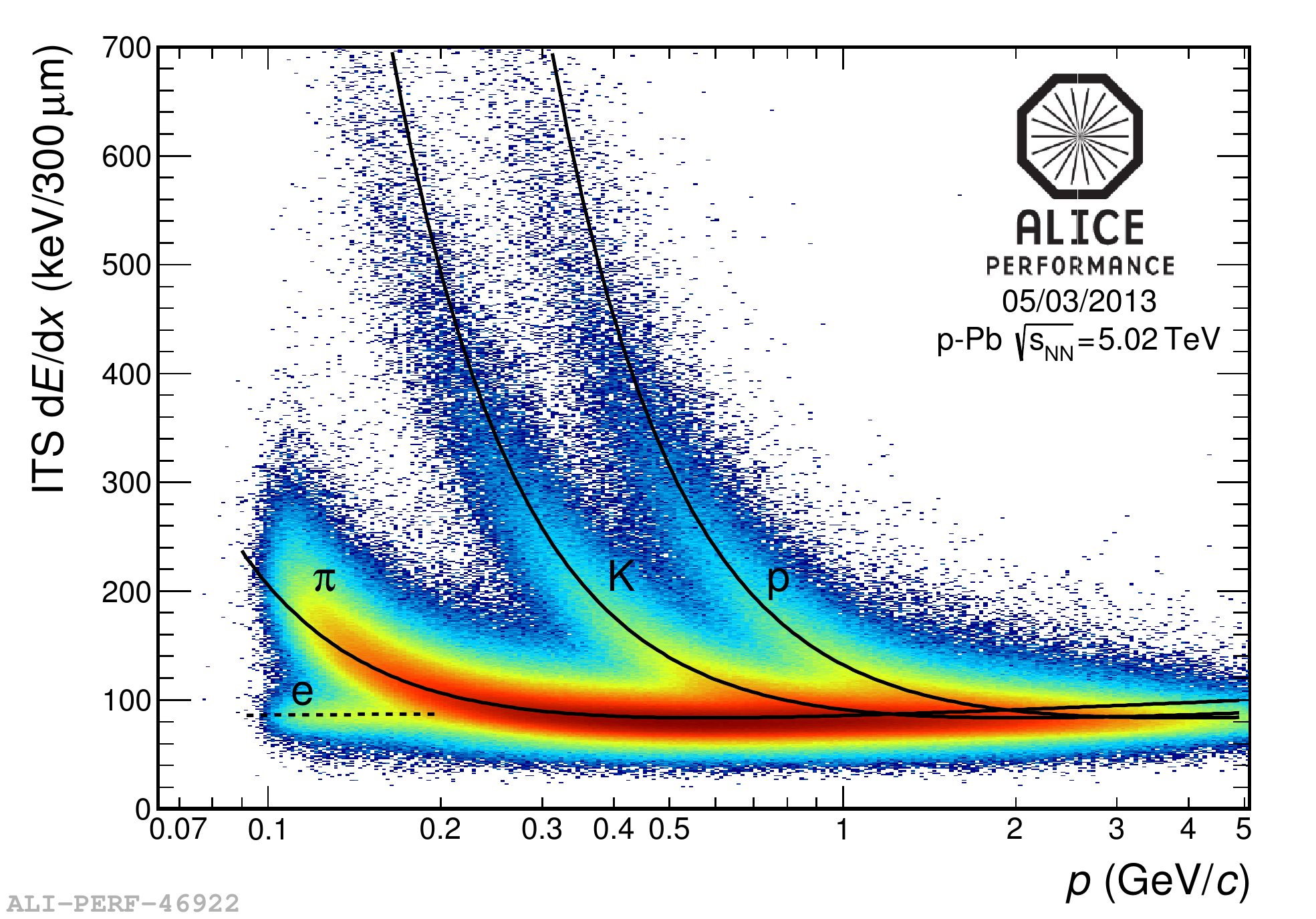}\figsubcap{d}}
  \caption{PID performance of barrel detectors: a) TOF signal versus particle momentum; b) TRD signal versus particle momentum; c) energy loses in TPC versus particle momentum; d) energy loses in ITS versus particle momentum.}%
  \label{fig:fig1}
\end{center}
\end{figure}

\section{Single and double diffraction}
\begin{figure}[h]
\begin{center}
\includegraphics[width=4.5in]{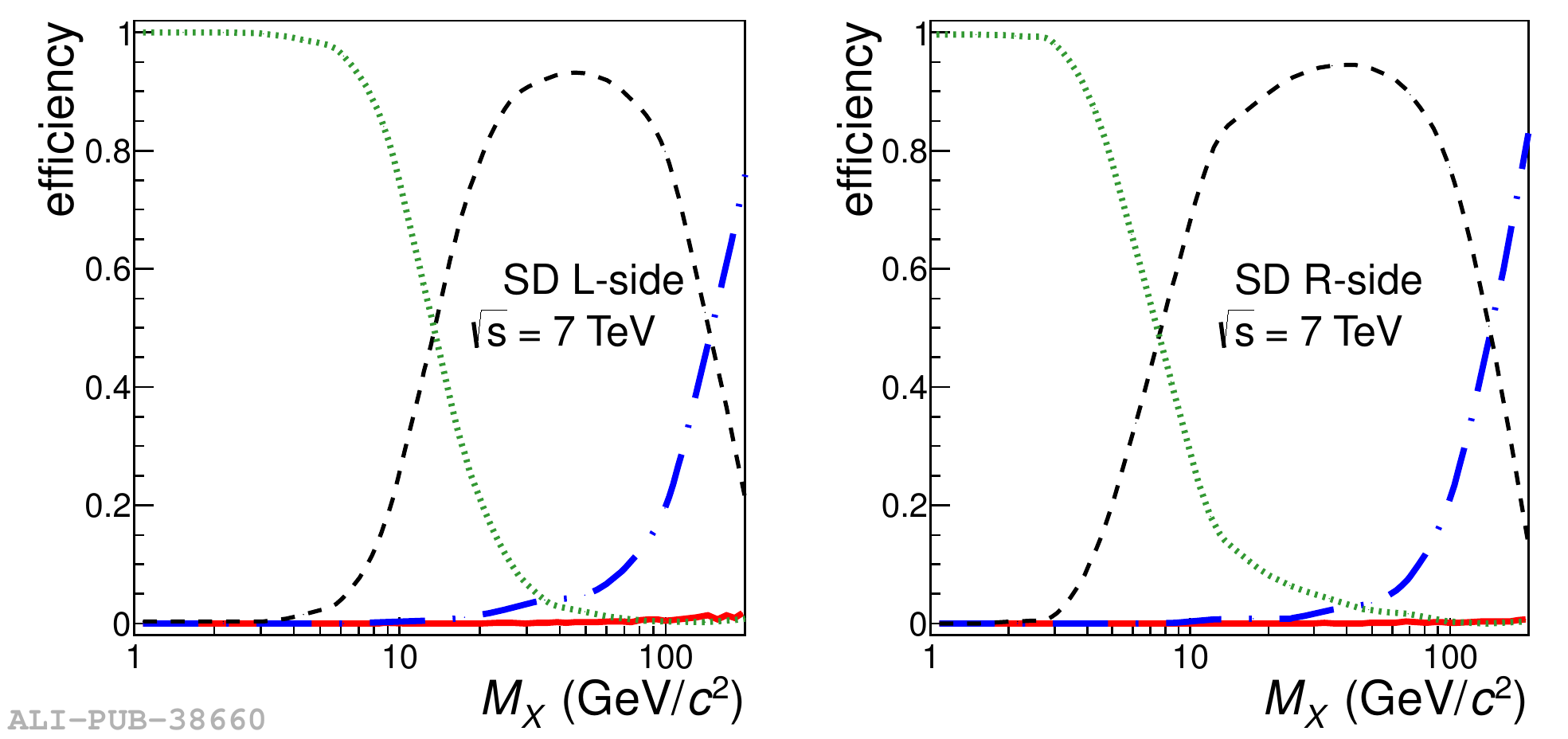}
\end{center}
\caption{Detection efficiency for single diffraction events for different diffractive cluster masses obtained with PYTHIA6 generator. L-side and R-side refer to the detector side at which SD occurred. {\it Green dotted line} shows detection inefficiency, {\it black dashed line} stands for detection efficiency of L(R)-side SD process as L(R)-side one-arm triggered event, {\it blue dashed-dotted line} shows detection efficiency as two-arm triggered event (misclassification), {\it red solid line} represents efficiency of L(R)-side SD process as R(L)-side one-arm triggered event (misclassification). }
\label{fig:fig2}
\end{figure}

The measurements of single and double diffraction cross sections\cite{analysis} were performed by classifying the events according to their topology. Namely, three offline triggers were developed:
\begin{itemlist}
\item Right-side one-arm trigger -- activity in the central or right-side (positive pseudorapidity) forward detectors, absence of any signals in the left-side (negative pseudorapidity) forward detectors;
\item Left-side one-arm trigger -- activity in the central or left-side forward detectors, absence of any signals in the right-side forward detectors;
\item Two-arm trigger -- activity in both left- and right-side forward detectors.
\end{itemlist}
The one-arm trigger events are assumed to be enriched by single-diffraction processes, while two-arm triggered events are very likely to originate from non-single diffraction. Fig.\ref{fig:fig2} shows the detection efficiency for single diffraction events for different diffractive cluster masses $M_X$. For small $M_X$ values, all produced particles have pseudorapidities close to that of the incoming proton and are not detected. With increasing $M_X$, the rapidity distribution broadens and events are mostly classified as one-arm triggers. At high diffractive masses $\sim$ 200 GeV/$c^2$, emitted particles cover the whole pseudorapidity range of the ALICE detector and therefore such events are classified as non-diffractive.   

\begin{figure}[h]%
\begin{center}
  \parbox{2.1in}{\includegraphics[width=2in]{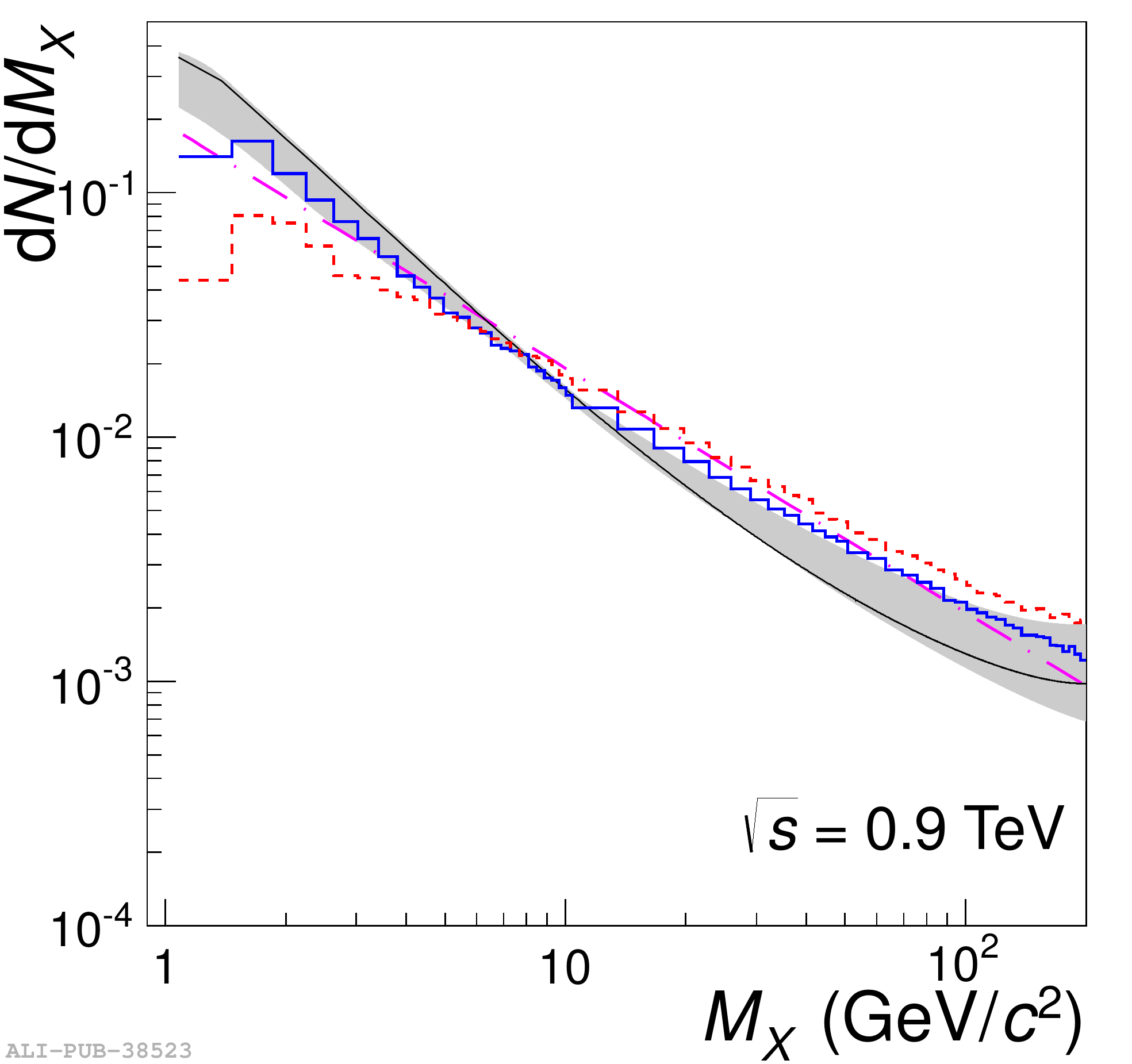}\figsubcap{a}}
  \hspace*{4pt}
  \parbox{2.1in}{\includegraphics[width=2in]{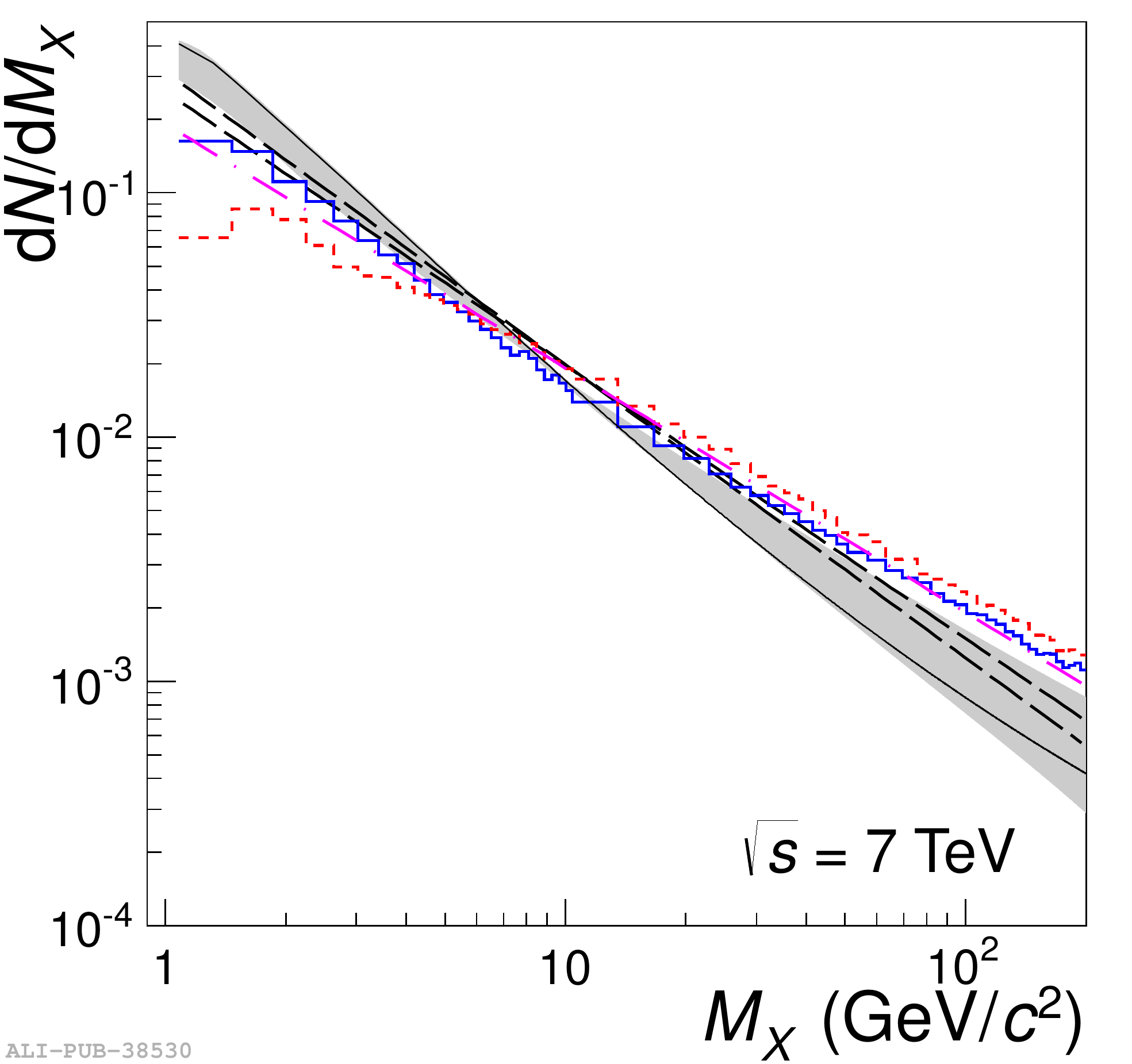}\figsubcap{b}}
  \caption{Diffractive-mass distributions, normalised to unity, for the SD process in pp collision at $\sqrt{s} = 0.9$ TeV (a) and $\sqrt{s} = 7$ TeV (b) from Monte Carlo generators PYTHIA6 ({\it blue histogram}), PHOJET ({\it red dashed-line histogram}), and Kaidalov-Poghosyan model\cite{KPmodel} ({\it black line}) -- used in present analysis. {\it Shaded area} is delimited by variation of the KP model, multiplying the distribution by a linear function which increases the probability at the threshold mass by a factor 1.5 (normalisation to unity has been kept) and by Donnachie-Landshoff parametrization\cite{DLpar}. {\it Magenta dotted-dashed line} represents $1/M_X$ distribution. At $\sqrt{s} = 7$ TeV (b) {\it black dashed lines} show $1/M_X^{1+2\Delta}$ with $\Delta = 0.085$ and $\Delta = 0.1 $ used with PYTHIA8 generator in the ATLAS measurement of inelastic cross section\cite{ATLASmes}.}%
  \label{fig:fig3}
\end{center}
\end{figure}

Detection inefficiency at low masses imposes certain challenges on cross section measurements. Since a significant part of the single-diffraction cross section is determined by the low-mass dissociation ($M_X \lesssim 10$ GeV/$c^2$), one has to choose carefully the theoretical model for expanding to low masses and estimating the fraction of missed events.  There are several models of diffractive processes on the market, which predict different diffractive mass distributions (see Fig.\ref{fig:fig3} for some of them).  
ALICE used the Kaidalov-Poghosyan (KP) model\cite{KPmodel} (shown as the black line in Fig.\ref{fig:fig3}) for determining the mean and the other models for accessing the uncertainty in the present analysis\cite{analysis}.

\begin{figure}[h]%
\begin{center}
  \parbox{2.3in}{\includegraphics[width=2.2in]{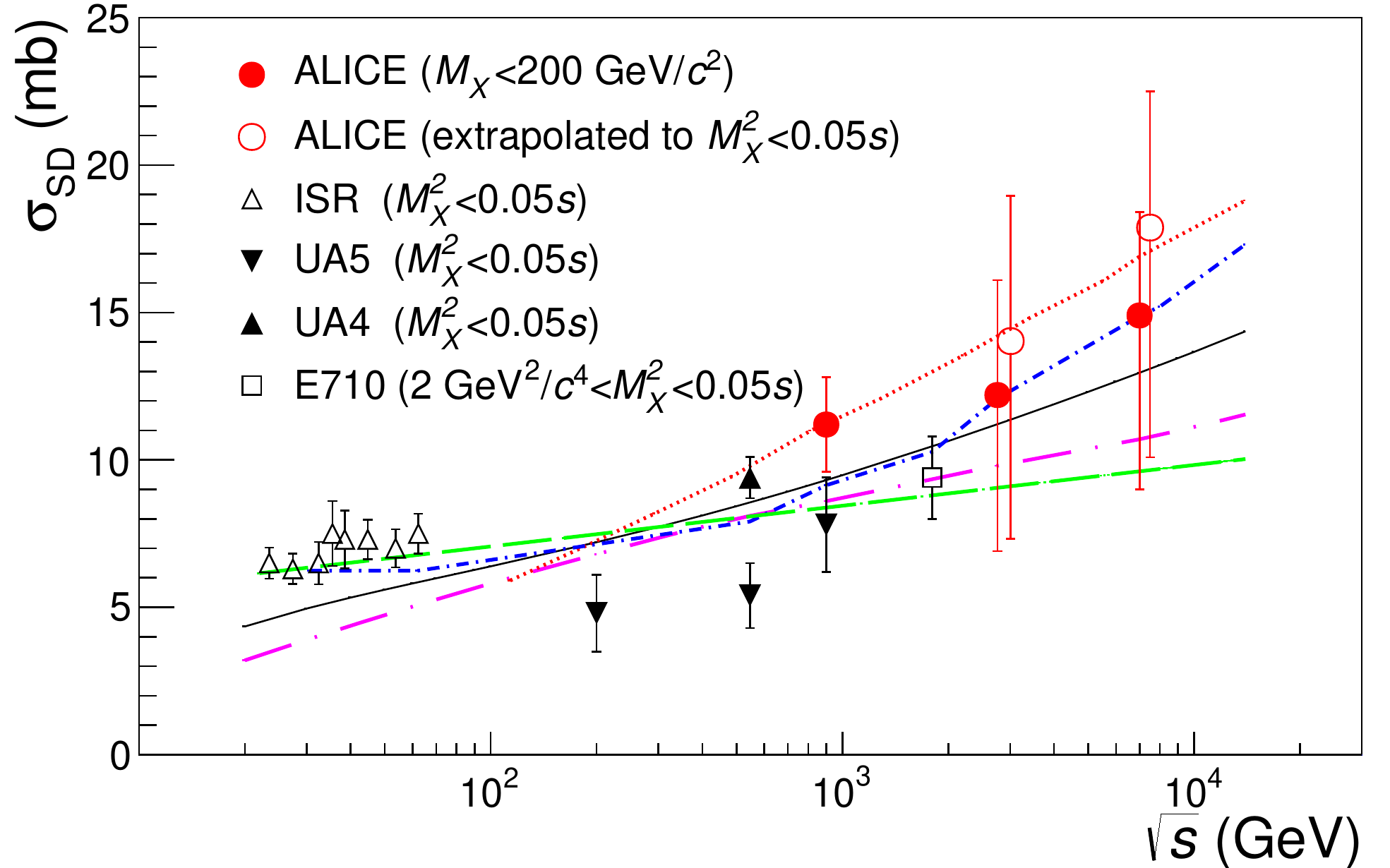}\figsubcap{a}}
  \hspace*{4pt}
  \parbox{2.3in}{\includegraphics[width=2.2in]{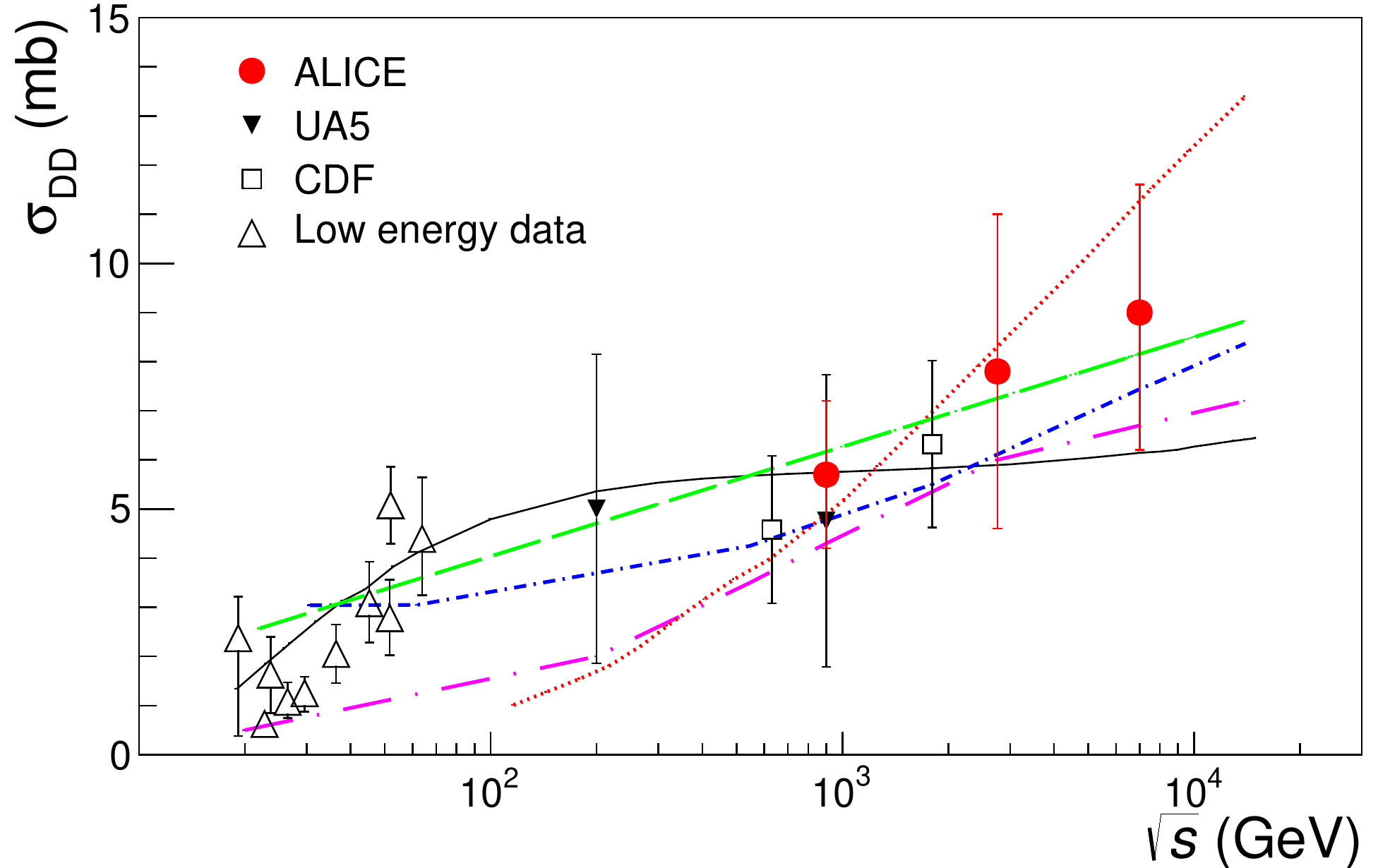}\figsubcap{b}}
  \caption{Single (a) and double (b) diffraction cross section measurements\cite{analysis} ({\it red solid circles}) comparing with different models ({\it short dot-dashed blue line}\cite{gotsman}, {\it dashed green line}\cite{goulianos}, {\it solid black line}\cite{KPmodel}, {\it long dot-dashed pink line}\cite{ostapchenko}, {\it dotted red line}\cite{ryskin}). (a) Data points from other experiments taken from Ref.~\citenum{59,60,61,62}. (b) Data points from other experiments taken from Ref.~\citenum{63}.
}%
  \label{fig:fig4}
\end{center}
\end{figure}

Monte Carlo generators PYTHIA6 (Perugia-0, tune 320)\cite{25,26,27} and PHOJET\cite{28} have been modified to follow the KP model for diffractive mass $M_X$ distribution and by tuning the single- and double-diffraction fractions in order to reproduce the rapidity gap distribution observed in two-arm triggered events and to match the measured ratios of one-arm and two-arm triggers. Trigger efficiencies have been evaluated for diffractive and non-diffractive events to extract SD and DD contributions to the inelastic cross section. As SD events with high diffractive masses $M_X$ end-up as two-arm triggers (see Fig.\ref{fig:fig2}), our mass limit for SD processes is $M_X < 200$~GeV/$c^2$. The double-diffraction fraction has been evaluated from adjusted Monte Carlo models as a fraction of events with $\Delta \eta > 3$ irrespectively of the generator subprocesses. The difference between two generators is attributed to a systematical uncertainty of the method. 
Finally, the fractions obtained have been transformed to the cross sections using the trigger cross section measured by ALICE via the van der Meer scan technique at $\sqrt{s} = 2.76$ TeV and $\sqrt{s} = 7$ TeV \cite{analysis}. As for the energy $\sqrt{s} = 0.9$ TeV, the inelastic cross section was evaluated by extrapolation\cite{UA5ext,analysis} based on the UA5 result\cite{UA5}. Fig.\ref{fig:fig4} shows measured single and double diffractive cross sections comparing to other data and theoretical models. One has to be careful when comparing data points from various experiments due to a possible different definition of single and double diffraction in different collaborations. For example, the CDF collaboration \cite{cdf} defines DD events as those with $\Delta \eta > 3$, as ALICE analysis does, but in addition subtracts the contribution from non-Pomeron exchanges.

\section{Central diffraction measurements}
\begin{figure}[h]
\begin{center}
\includegraphics[width=3.5in]{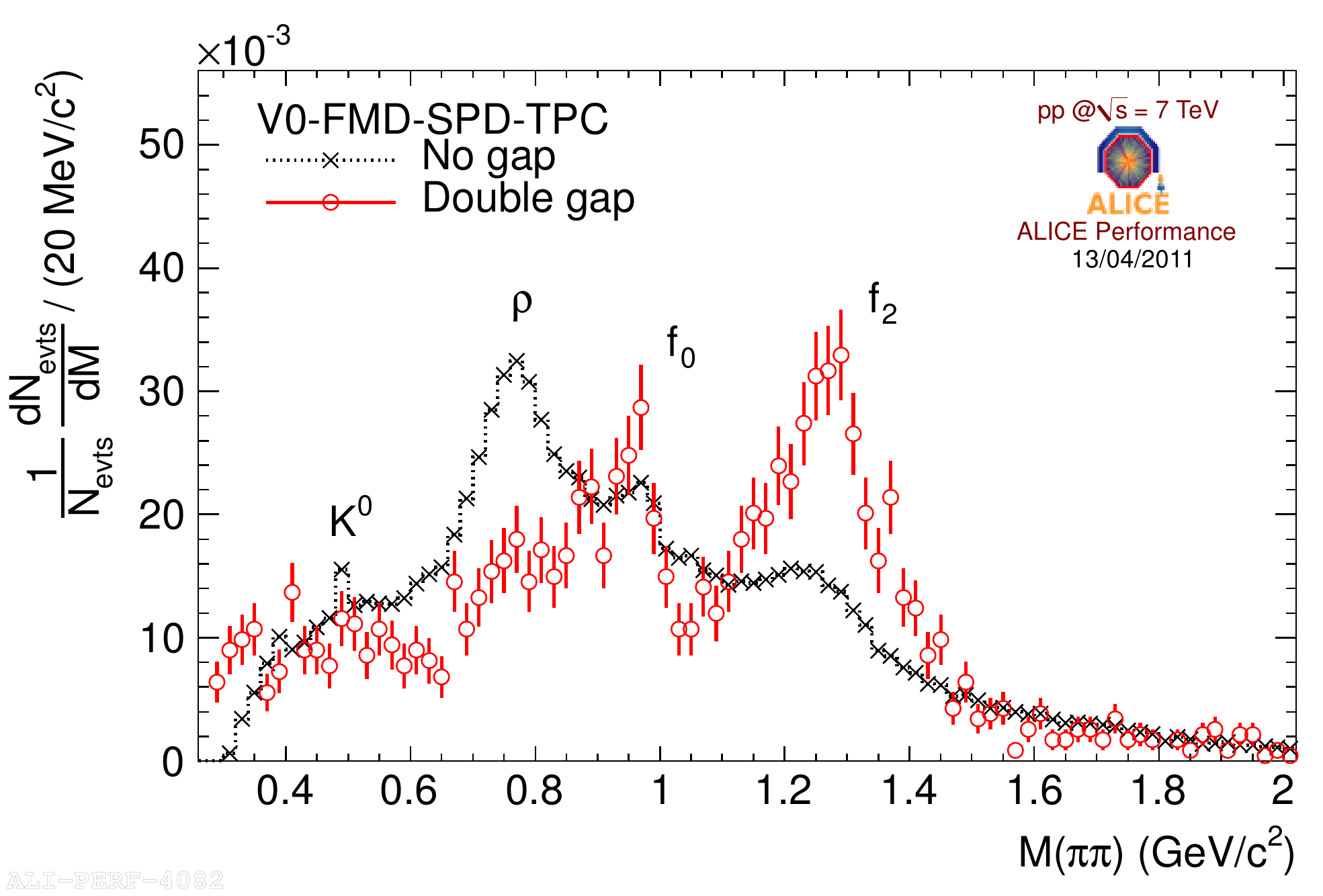}
\end{center}
\caption{Invariant mass distribution of events with two oppositely charged tracks assumed to be pions with and without double gap topology. Peaks are marked by most likely resonance candidates. }
\label{fig:fig5}
\end{figure}

The requirement of presence of an activity in the central barrel surrounded with large rapidity gaps from both sides can be used to select events enriched with the Pomeron-Pomeron fusion process. Since the outgoing protons are not detected in ALICE, three different processes may contribute to the double gap events: central diffraction without dissociation of outgoing projectiles (pure CD), with diffractive dissociation of one of the protons (CD+SD) or with dissociation of both protons (CD+DD). These three processes are indistinguishable in the ALICE setup . Fig.\ref{fig:fig5} shows the invariant mass distribution of events with two oppositely charged tracks with assigned pion mass with and without the double-gap topology. Peaks are marked by the resonance candidates. The centrally produced system in Pomeron-Pomeron fusion can have the quantum numbers with even angular moments and positive C and P parity, like $0^{++}, 2^{++}, 4^{++}$ and so on. The observed mass spectrum tends to follow this expectation but for the final conclusion a partial wave analysis is being performed by ALICE.

In the case of the central diffraction cross section measurements, the observed double gap event yield ratio to the minimum bias processes has been found\cite{felix} to be uniform over various data taking periods, and detectors had well performed during the Run1 of the LHC.
The next step is to correct the observed event fraction for the detector efficiency and define the cross section for double gap events.     

\section{Upgrade plans}
New scintillator counters called AD ({\bf A}lice {\bf D}iffractive) are planned to be installed in addition to the present experimental setup to extend the pseudorapidity coverage. The new detectors will be located $\sim$17 meters away from the interaction point on the left side of ALICE and  $\sim$19 meters away from the interaction point on the right side. They will cover the pseudorapidity ranges $-7.0 < \eta < -4.9$ on the left side and $4.7 < \eta < 6.4$ on the right side. Together with the other ALICE detectors, the AD will extend the acceptance for the low masses in single- and double-diffractive processes and increase the purity of double gap processes. Sensitivity to low diffractive masses and increased purity of double gap events helps to reduce the model dependence of diffractive measurements. The ALICE collaboration plans to start collecting data with these new detectors from the beginning of LHC Run2.

\section{Conclusion}
ALICE has measured the contributions of single- and double-diffractive
processes to the inelastic pp cross section. 
The cross sections of SD processes were obtained for diffractive
masses below 200 GeV/$c^2$. As for the DD processes, the cross
sections were obtained for events with a gap width $\Delta \eta > 3$. 
The obtained cross sections were compared with other measurements
at lower energies and with predictions from current models. They were
found to be consistent with all of them, within the obtained uncertainties.
In the case of central production, work is ongoing on the partial wave analysis of centrally produced two-mesons system as well as on the spectra and cross section measurements. 
The installation of the new scintillator counters -- AD -- will
increase the acceptance for low diffractive masses. The ALICE collaboration plans to start collecting data with these new counters from the
beginning of LHC Run2 in 2015.

\section*{Acknowledgements}
This work was partially supported by RFBR grant 12-02-91524.

\end{document}